\documentclass{PoS}

\usepackage{amsmath,amssymb}
\usepackage{calc}
\usepackage{enumitem}

\DeclareMathOperator{\sign}{sign}

\DeclareMathOperator{\myspan}{span}

\title{Double-pass variants for multi-shift BiCGstab($\ell$)}

\ShortTitle{Double-pass multi-shift BiCGstab($\ell$)}

\author{\speaker{Simon Heybrock}\thanks{Supported by the DFG collaborative research center SFB/TR-55 ``Hadron Physics from Lattice QCD''.}\\
        Institute for Theoretical Physics, University of Regensburg, 93040 Regensburg, Germany\\
        E-mail: \email{simon.heybrock@physik.uni-regensburg.de}}

\abstract{In analogy to Neuberger's double-pass algorithm for the Conjugate Gradient inversion with multi-shifts we introduce a double-pass variant for BiCGstab($\ell$). One possible application is the overlap operator of QCD at non-zero chemical potential, where the kernel of the sign function is non-Hermitian. The sign function can be replaced by a partial fraction expansion, requiring multi-shift inversions. We compare the performance of the new method with other available algorithms, namely partial fraction expansions with restarted FOM inversions and the Krylov-Ritz method using nested Krylov subspaces.}

\FullConference{The XXVIII International Symposium on Lattice Field Theory, Lattice2010\\
		June 14-19, 2010\\
		Villasimius, Italy}

\begin{document}

\section{Introduction and Motivation}
In this contribution we present double-pass variants for the multi-shift inverter BiCGstab($\ell$)\footnote{$\ell$ is the degree of the Minimal Residual polynomial in the algorithm} which, in some cases, can perform better than the conventional single-pass. The method is an analogue to Neuberger's double-pass Conjugate Gradient (CG) method \cite{Neuberger1999,Chiu2003}. The use of BiCGstab($\ell$) instead of CG can be a speed advantage (for Hermitian matrices) or necessary (for non-Hermitian matrices). One possible application is the computation of quark propagators for a set of distinct masses. Here, however, we focus on computing the overlap operator of QCD. At non-zero quark chemical potential, $\mu \ne 0$, it is defined as
\begin{align}
    D_{ov}(\mu) &= 1+\gamma_5 \sign \left( \gamma_5 D_w(\mu) \right),
\end{align}
where $D_w(\mu)$ is the (Wilson) Dirac operator with chemical potential. For $\mu\ne0$ the matrix $\gamma_5 D_w(\mu)$ is non-Hermitian. One way to compute the sign function of such a matrix, acting on a given vector $b$, is via a partial fraction expansion (PFE),
\begin{align}
    f(A)b &\approx \sum_{s=1}^{N_s} \frac{\omega_s}{A+\sigma_s}b,
    \label{eq:pfe}
\end{align}
where we are especially interested in the case of $A=(\gamma_5 D_w)^2$ with $f(A)=1/\sqrt{A}$, since $\sign z = z/\sqrt{z^2}$. The vectors $(A+\sigma_s)^{-1}b$ for a set of shifts $\{\sigma_s\}$ can be approximated by iterative inverters which find solutions in a Krylov subspace, defined as
\begin{align}
    \mathcal{K}_k(A,b) &= \myspan( b, Ab, \dotsc, A^{k-1}b ).
\end{align}
A crucial feature of Krylov subspaces is their shift invariance, $\mathcal{K}_k(A+\sigma_s,b) = \mathcal{K}_k(A,b)$, which allows for so called \emph{multi-shift inversions}, where one Krylov subspace suffices to compute $(A+\sigma_s)^{-1}b$ for a set $\{\sigma_s\}$ with little overhead per additional shift. We will refer to methods employing Eq. \eqref{eq:pfe} as PFE methods.

\section{Double-pass algorithm}
As a starting point, we consider established algorithms to compute the sign function of a non-Hermitian matrix, (i) the Krylov-Ritz method with nested Krylov subspaces, introduced in \cite{Bloch2009a}, (ii) PFEs with FOM inversions, introduced in \cite{Bloch2009} and (iii) PFEs with BiCGstab($\ell$) as inverter. The latter has so far not been considered in the context of the sign function. For details on the BiCGstab($\ell$) method see \cite{Sleijpen1998}, a version with shifts was introduced in \cite{Frommer2003}.
\begin{figure}
\begin{center}
    \includegraphics[height=0.7\textwidth,angle=-90]{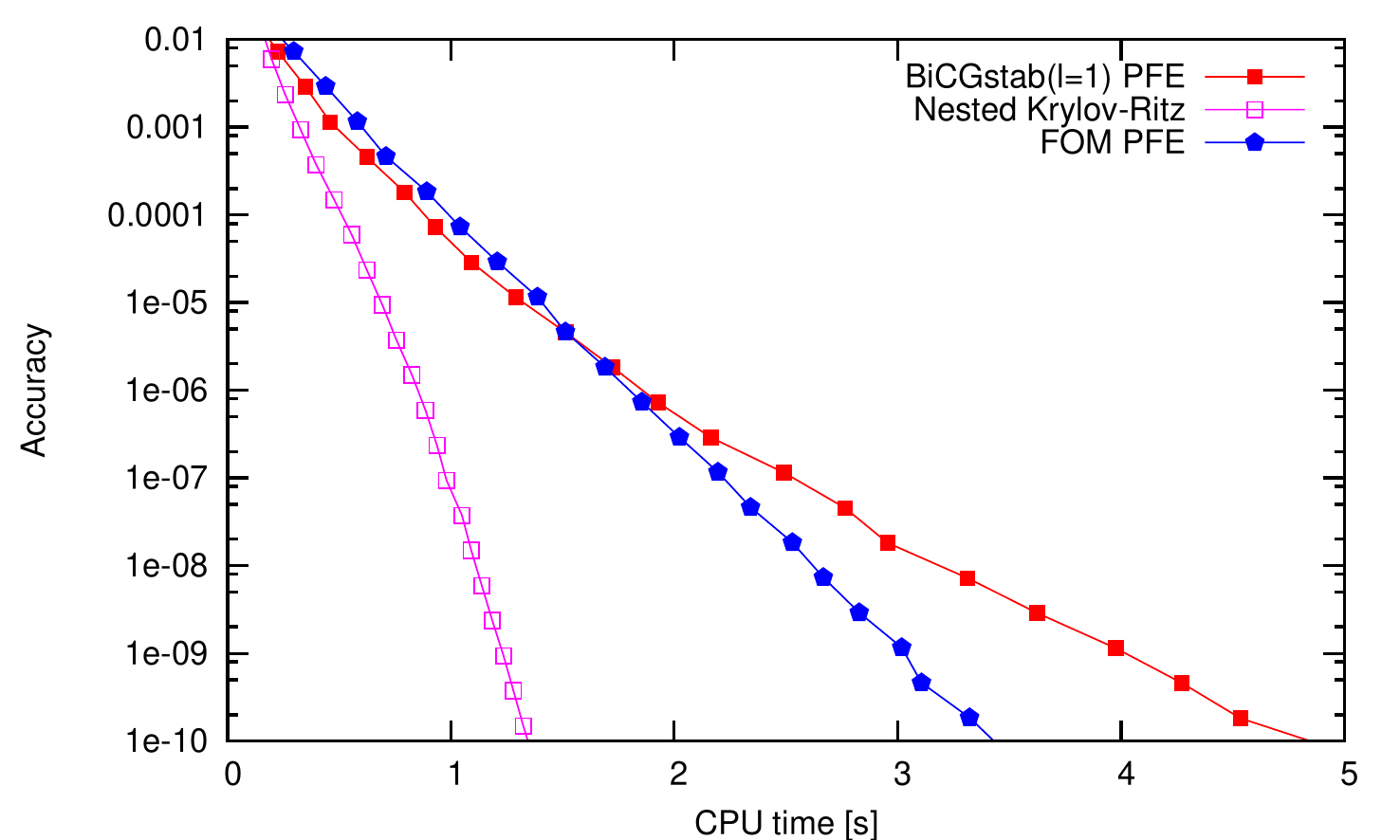}
    \caption{Accuracy vs. computation time for the overlap operator on a $4^3\times8$ lattice, $\beta=5.32$, $\mu=0.05$ with the Neuberger PFE. The number of poles $N_s$ is chosen minimal for the desired accuracy as in \cite{Bloch2009}. In this plot it ranges from $10$ (accuracy $0.01$) to $58$ (accuracy $10^{-10}$). The $4$ eigenvalues smallest in magnitude are deflated in advance.}
    \label{fig:established}
\end{center}
\end{figure}
Benchmark results are given in Fig. \ref{fig:established}. The nested Krylov-Ritz algorithm outperforms both PFE methods, which is somewhat surprising since all rely on a similar Krylov subspace.\footnote{Note however that the employed single-pass (nested) Krylov-Ritz method requires a huge amount of memory.} We can gain more insight by analysing the bad performance of BiCGstab($\ell$) in this case:
\begin{itemize}[itemsep=0mm,topsep=1mm]
    \item Denote by $N_s$ the number of shifts and by $M_s$ the number of outer iterations of the BiCGstab($\ell$) algorithm until the system with shift $\sigma_s$ is converged. Then, the multi-shift version has $\sum_{s=1}^{N_s} M_s l(0.5l+4.5)$ ``$axpy$'' operations ($y \leftarrow \alpha x + y$, for scalar $\alpha$ and vectors $x$ and $y$) more than the BiCGstab($\ell$) algorithm without shifts.
    \item BiCGstab($\ell$) requires $2l+5$ vectors and the multi-shift version has $N_s(l+1)$ additional shift vectors, where typically $N_s=\mathcal{O}(10)$. These figures should be seen in relation to the typical cache size of current processors ($\mathcal{O}(1~\mathrm{MByte})$) and the size of a vector, e.g., $50~\mathrm{kByte}$ (local volume $4^4$) or $800~\mathrm{kByte}$ (local volume $8^4$). In a typical case not all shift vectors fit into cache and the access to main memory can become the bottleneck of the algorithm.
\end{itemize}
To tackle these performance restraints one can try a \emph{double-pass} approach in analogy to Neuberger's double-pass algorithm for a multi-shift CG inversion. Schematically the idea is as follows: The quantity computed in Eq. \eqref{eq:pfe} and approximated in a Krylov subspace is
\begin{align}
        \sum_{s=1}^{N_s}\omega_s (A+\sigma_s)^{-1} b &\approx \sum_{s=1}^{N_s} \omega_s \sum_{n=1}^N w_s^{(n)},
        \label{eq:recursion}
\end{align}
where $N$ is the number of iterations in the inverter and $w_s^{(n)}$ is a vector for shift $s$ in iteration $n$. To remove $s$ dependent vectors one could try to swap the sums over $s$ and $n$, however $w_s^{(n)}$ is given by a recursion relation,
\begin{align}
        w_s^{(n)} &= \alpha_s^{(n)} w_s^{(n-1)} + \beta_s^{(n)} v^{(n)} = \sum_{i=1}^{n} \gamma_{s,i}^{(n)} v^{(i)},
        \label{eq:recursion_solved}
\end{align}
where $v^{(n)}$ is an unshifted iteration vector. In the last step the recursion of the vectors $w_s^{(n)}$ was resolved. By combining Eqs. \eqref{eq:recursion} and \eqref{eq:recursion_solved} and summing over $s$ (and $n$), all vectors depending on $s$ are removed from the algorithm. However, the coefficients $\gamma_i = \sum_{s,n}\gamma_{s,i}^{(n)}$ are not known until the end of the iteration. There are two options
\begin{enumerate}[itemsep=0mm,topsep=1mm]
    \item (\emph{double-pass}): Follow Neuberger's approach by running the algorithm once to obtain $\gamma_i$. In a second pass generate the vectors $v^{(i)}$ again and compute $\sum_i \gamma_i v^{(i)}$.
    \item ($\emph{pseudo-double-pass}$): Compute the coefficients $\gamma_i$ as in double-pass, but store all $v^{(i)}$ during the first pass instead of recomputing them in a second pass.
\end{enumerate}
Both methods remove all $s$-dependent vectors from the algorithm, hence reducing the number of operations and number of vectors to be held in cache. In our case, to obtain the coefficients corresponding to the (schematic) coefficients $\gamma_{i}$, the recursion has to be solved for the BiCGstab($\ell$) algorithm. The result is given in Sec. \ref{algorithm}.

\section{Cost analysis and benchmarks}
The number of operations (scalar ones are omitted) and vectors of the multi-shift BiCGstab($\ell$) algorithms are given in the following table ($M$ denotes the number of outer iterations of the algorithm and the dimension of the Krylov space is $2Ml$):\vspace{2mm}\\
\begin{small}
\begin{tabular}{l||l|l|l||l}
Method                    &  \#Mv & \#axpy                                            & \#dot-products  & \#vectors           \\
\hline
1-pass               &  $2Ml$ & $Ml(1.5l+5.5)+\sum_{s=1}^{N_s} M_s l(0.5l+4.5)$ & $Ml(0.5l+3.5)$ &  $2l+5+N_s(l+1)$ \\
2-pass               &  $4Ml$ & $Ml(1.5l+5.5)+Ml(1.5l+4.5)+2Ml$               & $Ml(0.5l+3.5)$  &  $2l+5$     \\
pseudo-2-pass        &  $2Ml$ & $Ml(1.5l+5.5)+2Ml$                           & $Ml(0.5l+3.5)$ &  $2l+5+2Ml$   \\
\end{tabular}\vspace{2mm}\\
\end{small}
The number of vectors alone is not always meaningful: In single-pass a considerable subset\footnote{depending on $\ell$, and on the removal of converged systems from the iteration} of the $N_s(l+1)$ vectors is accessed in each iteration of the algorithm. In pseudo-double-pass each of the $2Ml$ vectors is written and read exactly once, in total. That is, the access pattern of pseudo-double-pass requires less memory access than single-pass, even though many more vectors are involved.

\begin{figure}
\begin{center}
\includegraphics[height=0.7\textwidth,angle=-90]{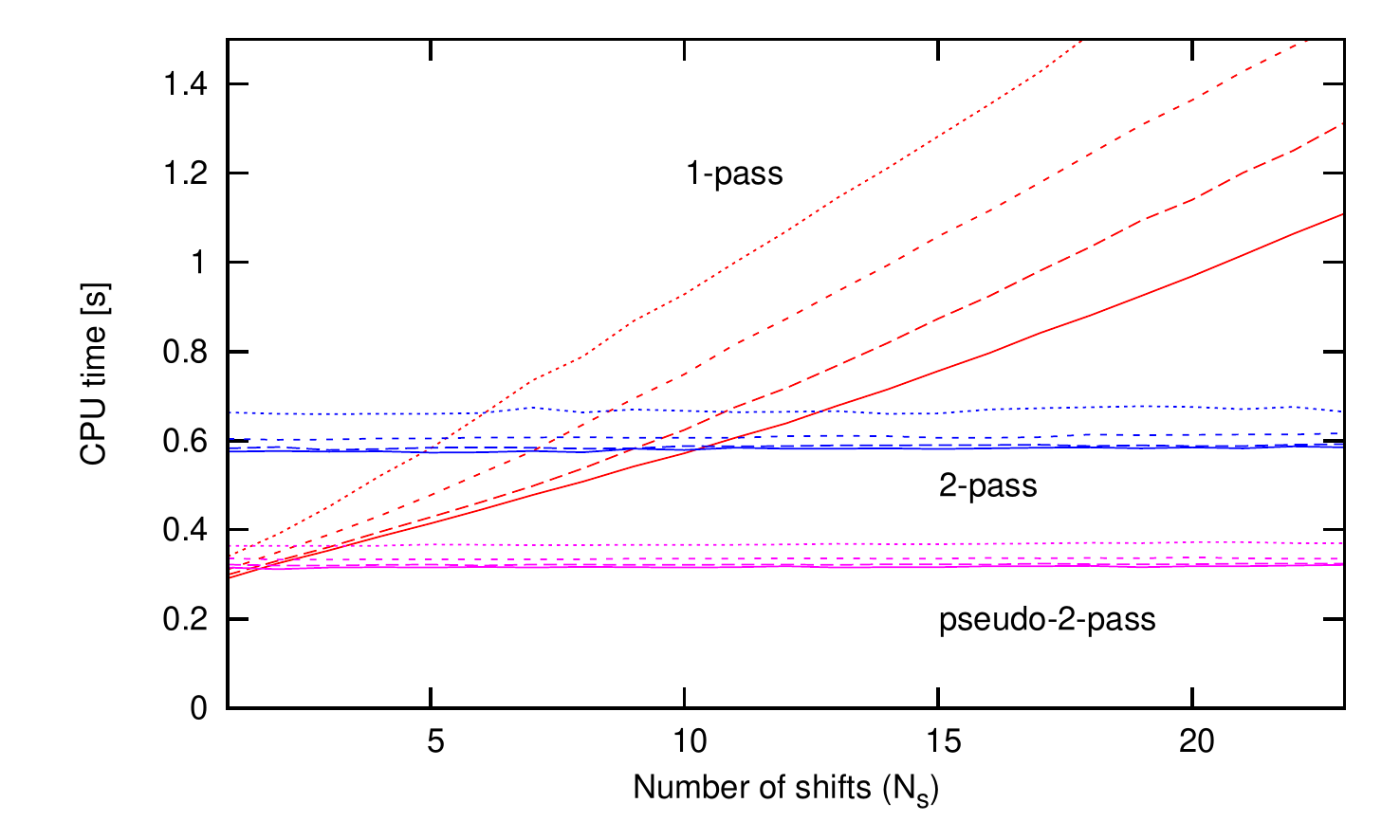}
\caption{Computation time vs. $N_s$ for fixed $Ml=256$ for a $4^3\times8$ lattice (Wilson Dirac operator) for all three BiCGstab($\ell$) variants. Results are given for $l=1,2,4,8$ with lines solid to dotted.}
\label{fig:scaling}
\end{center}
\end{figure}
As a naive test of the figures given in the table we consider the algorithm runtime for fixed $Ml$ with a varying number of shifts, given in Fig. \ref{fig:scaling}. The two-pass and pseudo-two-pass runtime is largely independent of $N_s$. The pure operation count of two-pass would yield an almost doubled computation time compared to pseudo-two-pass. In practice, however, it is less since no (or less) main memory access is required. An effect of the cache size can be seen from the single-pass $l=1$ curve. The slope changes in the vicinity of $N_s=10$, which is consistent with the cache size of $4~\mathrm{MByte}$ and the size of a vector of $100~\mathrm{kByte}$. As a further observation, the relative performance loss for large $\ell$ is much smaller in the double-pass methods compared to single-pass, which is not surprising since the number of required vectors increases with $\ell$.\footnote{Since we work at fixed $Ml$ this plot does not tell which $\ell$ is optimal, since the convergence rate depends on $\ell$.}

\begin{figure}
\begin{center}
    \includegraphics[height=0.7\textwidth,angle=-90]{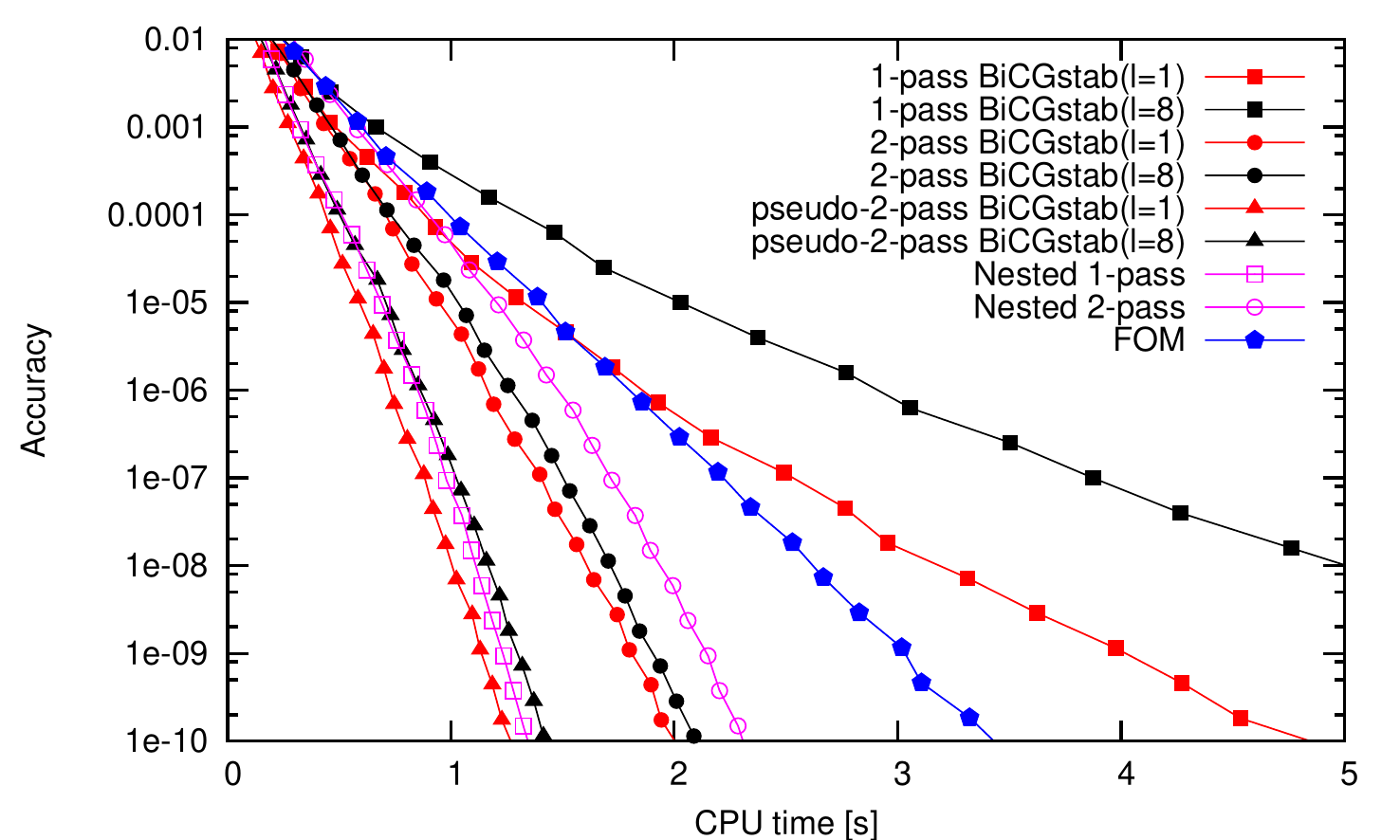}
    \caption{Accuracy vs. computation time for the overlap operator ($4^3\times8$ lattice with $\beta=5.32$, $\mu=0.05$). The timings are averaged over $200$ independent gauge configurations. Note that an extreme case with little deflation ($4$ eigenvalues smallest in magnitude) was chosen for this plot where many poles are required (as before $N_s$ is scaled from $10$ to $58$), yielding a large speed advantage of the double-pass algorithms. When less poles are needed (e.g., for small $\mu$ when the Zolotarev PFE can be used instead of the Neuberger expansion) the performance difference is often smaller.}
    \label{fig:convergence}
\end{center}
\end{figure}
As a more realistic benchmark we compute the overlap operator for given configurations in Fig. \ref{fig:convergence}. The double-pass and pseudo-double-pass BiCGstab($\ell$) algorithms perform as well or even better than the nested double-pass and single-pass algorithms, respectively. Note that also the respective memory requirements are similar. In double-pass the performance does not degrade for $l>1$ as it does for single pass. To explore differences between the nested Krylov-Ritz method and the BiCGstab($\ell$) methods a series of benchmarks was performed, where both the number of deflated eigenvectors and the chemical potential $\mu$ were varied. The tests indicate that BiCGstab($\ell$) profits more from deflation than the Krylov-Ritz method does. On the other hand, for large $\mu$, BiCGstab($\ell$) tends to stagnate earlier than Krylov-Ritz.

\begin{figure}
\begin{center}
    \includegraphics[height=0.55\textwidth,angle=-90]{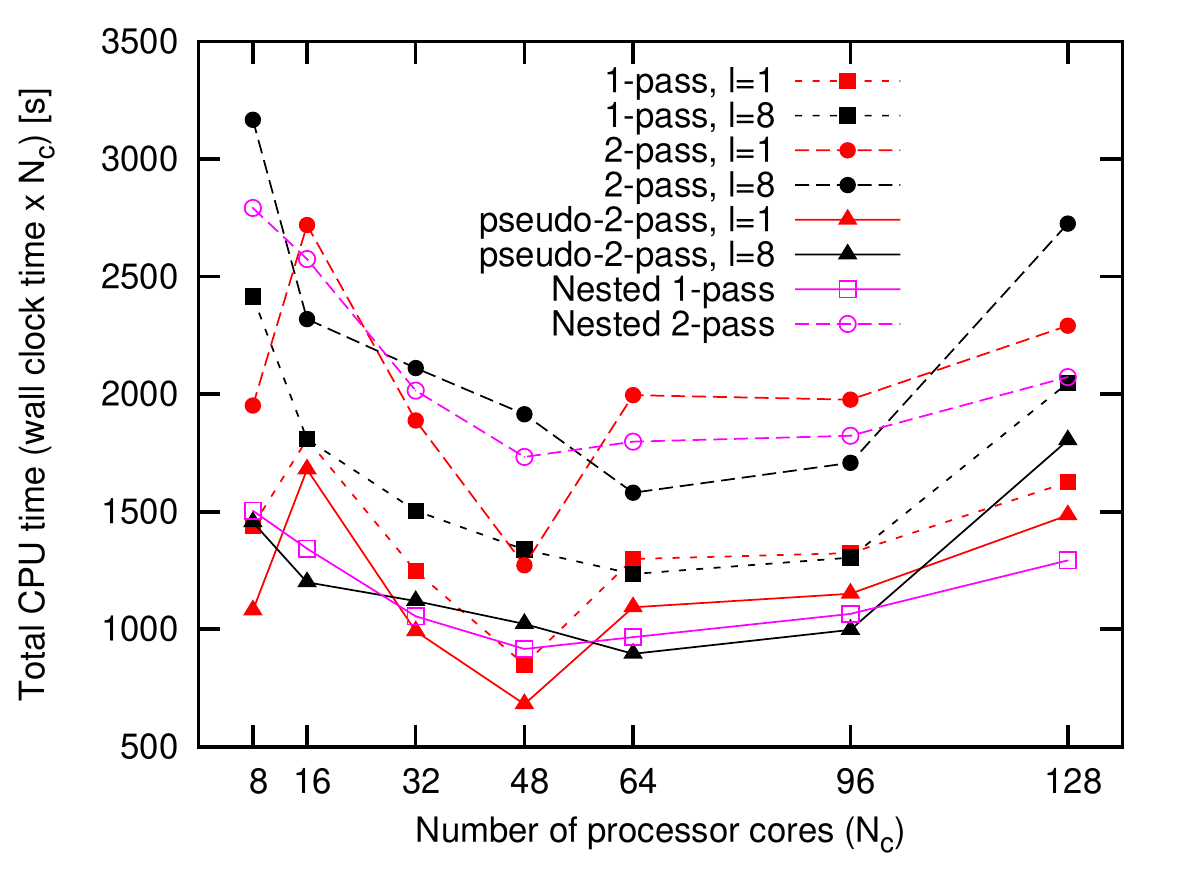}
    \caption{Total computation time vs. number of cores for the nested Krylov-Ritz method and pseudo-double-pass BiCGstab($\ell$) for $l=1$ and $8$. The simulation uses a $12^3 \times 24$ lattice, $\beta=5.71$, $\mu=0.016667$ with $44$ deflated eigenvalues. The accuracy is $10^{-10}$, obtained with $N_s=16$ shifts. The dimension of the Krylov subspace is about $2000$ for BiCGstab($\ell=1,8$) as well as for nested Krylov-Ritz. The benchmarks where done on an Opteron 2354 Cluster ($2.2~\mathrm{GHz}$, 2 quad-core processors per node, $16~\mathrm{GByte}$ RAM per node, Infiniband network). Lines are drawn to guide the eye.} 
    \label{fig:core_scaling}
\end{center}
\end{figure}
Finally we give results of a larger-scale simulation in Fig. \ref{fig:core_scaling}. As before, pseudo-double-pass BiCGstab($\ell$) is the algorithm performing best. The optimal $\ell$ depends on the number of cores $N_c$. Due to memory limitations for small $N_c$ and network limitations for large $N_c$, there is an optimal $N_c$ minimizing the total CPU time.

\section{Algorithm details}
\label{algorithm}
We follow the notation in Ref. \cite{Frommer2003} where also a listing of BiCGstab($\ell$) is given. Upper indices $mj$ denote iteration $j$ of BiCG part and outer iteration $m$. Define the coefficients
\begin{small}
\begin{align}
        A_{mj} &= \sum_{s} \omega_s  \frac{1}{(\vartheta^s \varphi^s)^{mj}} \sum_{j'=j}^{l-1} (\alpha^s)^{mj'}\left( \prod_{k=j+1}^{j'}(-\beta^s)^{mk} \right),\\
        B_{m}^s &= \sum_{n=m+1}^{M-1} \left\{ \sum_{j=0}^{l-1} (\alpha^s)^{nj} \left( \prod_{k=0}^{j} (-\beta^s)^{nk} \right) \right\} \left\{ \prod_{k=m+1}^{n-1}\left( \sum_{j=0}^l \frac{-\gamma_{j}^{\,k}}{(\psi^s)^k} \sigma_s^{j}(-1)^{l-j} \right)\prod_{k'=0}^{l-1}(\beta^s)^{kk'} \right\},\\
        D_{mij}^s &= \frac{-\gamma_i^{\,m}}{(\psi^s)^m}\frac{1}{(\vartheta^s \varphi^s)^{mj}}  \prod_{k=j+1}^{l-1} (-\beta^s)^{mk} ,\qquad
        E_{mj}^s = \frac{-1}{(\psi^s)^m} \left( \sum_{i=j+1}^l \gamma_i^{\,m} \sigma_s^{i-j-1} (-1)^{l-i} \right)  \prod_{k=j+1}^{l-1} (\beta^s)^{mk},\\
        F_{mj}^s &= \frac{1-(\alpha^s)^{mj}\sigma_s}{(\alpha^s)^{mj}(\vartheta^s\varphi^s)^{mj}},\qquad
        G_{mj}^s = -\frac{1}{(\alpha^s)^{mj}(\vartheta^s\varphi^s_{\mathrm{new}})^{mj}},
\end{align}
\end{small}
where $\gamma_0=-1$, and a matrix
\begin{small}
\begin{align}
        (M^m)_{jk} &= - \sum_{q=k}^{j-1} \alpha^{mq} \left( \prod_{p=k+1}^q (-\beta^{mp}) \right),\qquad j,k=0,\dotsc,l-1.
\end{align}
\end{small}
Then the contribution to $\sum_s \omega_s x^s$ from the BiCG part is given by
\begin{small}
\begin{align}
    \mathbf{x}_{\mathrm{BiCG}} &= \sum_{m=0}^{M-1} \sum_{j=0}^{l-1} \left\{ \sum_{p=j}^{l-1} \sum_{k=j}^p \left[ A_{mp} ( (M^m)^j)_{pk} + \sum_{i=0}^j \left( \sum_s \omega_s B_m^sD_{mip}^s \right) ( (M^m)^{j-i})_{pk} \right] \right.\notag\\
    &\qquad\times\left.\left[ (\mathbf{r}_j)^{mj}-\sum_{q=j}^{k-1} \alpha^{mq} \left( \prod_{p'=j+1}^q (-\beta^{mp'}) \right)(\mathbf{u}_{j+1})^{mj} \right] \right.\notag\\
        &\quad+ \left. \left( \sum_s \omega_s B_m^s E_{mj}^s F_{mj}^s \right) (\mathbf{r}_j)^{mj} +  \left( \sum_s \omega_s B_m^s E_{mj}^s G_{mj}^s \right) \left( (\mathbf{r}_j)^{mj} -\alpha^{mj}(\mathbf{u}_{j+1})^{mj} \right)\right\}.
\end{align}
\end{small}
All operations involving the vectors $\mathbf{u}_i^s$ can be removed from the original algorithm. The final result is given by adding $\mathbf{x}_{\mathrm{BiCG}}$ to the contributions of the seed system and the MR-part of the algorithm, $\sum_s \omega_s \mathbf{x}_{MR}^s$, which is computed trivially. For reference an implementation of the algorithms is provided online at \href{http://sourceforge.net/projects/bicgstabell2p/}{http://sourceforge.net/projects/bicgstabell2p/}.

\section{Conclusions}
We have presented an extension of the double-pass trick from Conjugate Gradient to the more general BiCGstab($\ell$). While initially PFE methods looked inferior to the nested Krylov-Ritz method in the non-Hermitian case, our new (pseudo-)double-pass BiCGstab($\ell$) is a method with similar performance. Our benchmarks concentrated on the overlap operator where pseudo-double-pass performs as well or even better than the nested Krylov-Ritz method on the tested architectures. Current supercomputers might have enough main memory such that pseudo-double-pass is feasible, but this will depend on details of the simulation. Large values of $\ell$ yield less overhead in the double-pass methods compared to single-pass.
This could boost the application of the algorithm in problems where $l>1$ is crucial for convergence.
 We plan to investigate the efficiency of the double-pass BiCGstab($\ell$) algorithms for other functions aside from the inverse square root.

As a closing remark let us mention that a pseudo-double-pass method can also be used instead of the usual (double-pass) Conjugate Gradient method. This extension seems trivial, though we are not aware of any mention in the literature.

\section*{Acknowledgements}
I want to thank Jacques C.R. Bloch and Tilo Wettig for support, advice and discussions.

\bibliographystyle{JHEP_JB}
\bibliography{/home/hes10653/data/doc/refs/generic}

\providecommand{\href}[2]{#2}\begingroup\raggedright\begin{thebibliography}{1}

\bibitem{Neuberger1999}
H.~Neuberger, {\em Minimizing storage in implementations of the overlap
  lattice-Dirac operator},  {\em Int. J. Mod. Phys. C} {\bf 10} (1999)
  1051--1058, [\href{http://xxx.lanl.gov/abs/hep-lat/9811019}{{\tt
  hep-lat/9811019}}].

\bibitem{Chiu2003}
T.-W. Chiu and T.-H. Hsieh, {\em A note on Neuberger's double pass algorithm},
  {\em Phys. Rev. E} {\bf 68} (2003) 066704,
  [\href{http://xxx.lanl.gov/abs/hep-lat/0306025}{{\tt hep-lat/0306025}}].

\bibitem{Bloch2009a}
J.~C.~R. Bloch and S.~Heybrock, {\em A nested Krylov subspace method to compute
  the sign function of large complex matrices},  {\em Comput. Phys. Commun.}
  (to be published) [\href{http://xxx.lanl.gov/abs/0912.4457}{{\tt
  arXiv:0912.4457}}].

\bibitem{Bloch2009}
J.~C.~R. Bloch {\em et.~al.}, {\em {Short-recurrence Krylov subspace methods
  for the overlap Dirac operator at nonzero chemical potential}},  {\em Comput.
  Phys. Commun.} {\bf 181} (Oct., 2010) 1378--1387,
  [\href{http://xxx.lanl.gov/abs/0910.1048}{{\tt arXiv:0910.1048}}].

\bibitem{Sleijpen1998}
G.~L.~G. Sleijpen and D.~R. Fokkema, {\em BiCGstab($\ell$) For Linear Equations
  Involving Unsymmetric Matrices With Complex Spectrum},  {\em Electronic
  Transactions on Numerical Analysis} {\bf 1} (1993) 11--32.

\bibitem{Frommer2003}
A.~Frommer, {\em BiCGStab($\ell$) for families of shifted linear systems},
  {\em Computing} {\bf 70} (2003), no.~2 87--109.

\end{thebibliography}\endgroup

%

\end{document}